\documentclass{amsart}
\usepackage{times}
\usepackage{amsthm}
\newtheorem{theorem}{Theorem}
\newtheorem{definition}{Definition}[theorem]
\usepackage{amssymb}
\usepackage[]{graphicx}
\newcommand{\K}{{\mathbb C}}
\newcommand{\C}{{\mathbb C}}
\newcommand{\R}{{\mathbb R}}

\renewcommand{\P}{{\mathbb P}}
\DeclareMathOperator{\argmin}{argmin}
\DeclareMathOperator{\tr}{tr}
\newcommand{\HR}[1]{{{#1}}}
\newcommand{\FK}[1]{{{#1}}}
\newcommand{\KK}{{\HR{L}}}
\newcommand{\MM}{{\HR{m}}}
\begin{document}



\title[Structured random measurements]{Structured random measurements in signal processing}


\authors{Felix Krahmer\footnote{e-mail: {\sf f.krahmer@math.uni-goettingen.de}, Phone: +49\,551\,39\,10584,
     Fax: +49\,551\,39\,33944}} \address{Institute for Numerical and Applied Mathematics, University of G\"ottingen, Lotzestr. 16-18, 37083 G\"ottingen, Germany}
     and Holger Rauhut\footnote{e-mail: {\sf rauhut@mathc.rwth-aachen.de}, Phone: +49\,241\,80\,94540, Fax: +49\,241\,80\,92390}
\address{Lehrstuhl C f{\"u}r Mathematik (Analysis), RWTH Aachen University, Templergraben 55, 52062 Aachen, Germany}
\begin{abstract}
Compressed sensing and its extensions have recently triggered interest in randomized signal acquisition. A key finding is that random measurements provide sparse signal reconstruction guarantees for efficient and stable algorithms with a minimal number of samples. While this was first shown for (unstructured) Gaussian random measurement matrices, 
applications require certain structure of the measurements leading to structured random measurement matrices.
Near optimal recovery guarantees for such structured measurements have been developed 
over the past years in a variety of contexts. This article surveys the theory in three scenarios: compressed sensing (sparse recovery), low rank matrix recovery, and phaseless estimation. The random measurement matrices to be considered include random partial Fourier matrices, partial random circulant matrices (subsampled convolutions), matrix completion, and phase estimation from magnitudes of Fourier type measurements. The article concludes with a brief discussion of the mathematical techniques for the analysis of such structured random measurements.
\end{abstract}
\maketitle             

\section{Introduction}
In the theory of inverse problems, structural properties of signals and images have always played an important role. Namely, most inverse problems arising in practical applications are ill-posed, which makes them impossible to solve in a robust manner without imposing additional assumptions. The expected or observed structure of the signal can then serve as a regularizer necessary to allow for efficient solution methods. 
At the same time, it was well-known that the success of such approaches heavily depended on the nature of the observed measurements. Usually, these measurements were considered to be given by the application, and the goal was to formulate properties that allow for successful reconstruction.

A different perspective was taken in a number of works over the last decade, starting with the seminal works of Donoho \cite{do06-2} and Cand{\`e}s, Romberg, and Tao \cite{carota06}. 
Namely, the goal was to use the degrees of freedom of the underlying applications to design measurement systems that by construction are well-suited for successful reconstruction of
structured signals. In many cases, as it turned out, measurements selected at random were shown to lead to superior, often near-optimal recovery guarantees. The tightest recovery
guarantees are typically obtained when structural constraints on the measurements as prescribed by the application are ignored and the measurement parameters are chosen completely
at random, for example following independent normal distributions. In a next step, the constraints then need to be reintroduced, resulting in structured random measurements. Error
analysis and recovery guarantees for such structured random measurement systems shall be the main focus of this survey 
article. We will focus on three types of signal recovery problems: compressed sensing, low rank matrix recovery, and phaseless estimation.

{\em Compressed sensing} is concerned with the recovery of approximately sparse signals from linear measurements. A signal is said to be {\em $k$-sparse} in a given basis or frame, if it can be expressed as a linear combination of only $k$ of the basis or frame elements. Approximate sparsity is a common model in signal and image processing, as natural signals are observed to be extremely well represented by restricting to just the very few largest representation coefficients in a suitable basis or frame and setting the remaining ones to zero. In fact, lossy compression schemes including JPEG, MPEG or MP3 are based on this empirical finding. Suitable representation systems include wavelet bases and shearlet frames, and approximate sparsity is also observed for the discrete gradient (though it does not constitute a basis or frame representation). Motivating applications for this problem setup include magnetic resonance imaging (MRI) \cite{dolupa07,albahalupava10}, coded 
aperture imaging \cite{mawi08}, remote sensing \cite{hest09,fastya09,hurast12}, and infrared imaging \cite{badadukelatati08}.

\begin{figure}\label{fig:SparseFourier}
\begin{center}
\parbox[t]{0.48\textwidth}{
\includegraphics[width=0.48\textwidth]{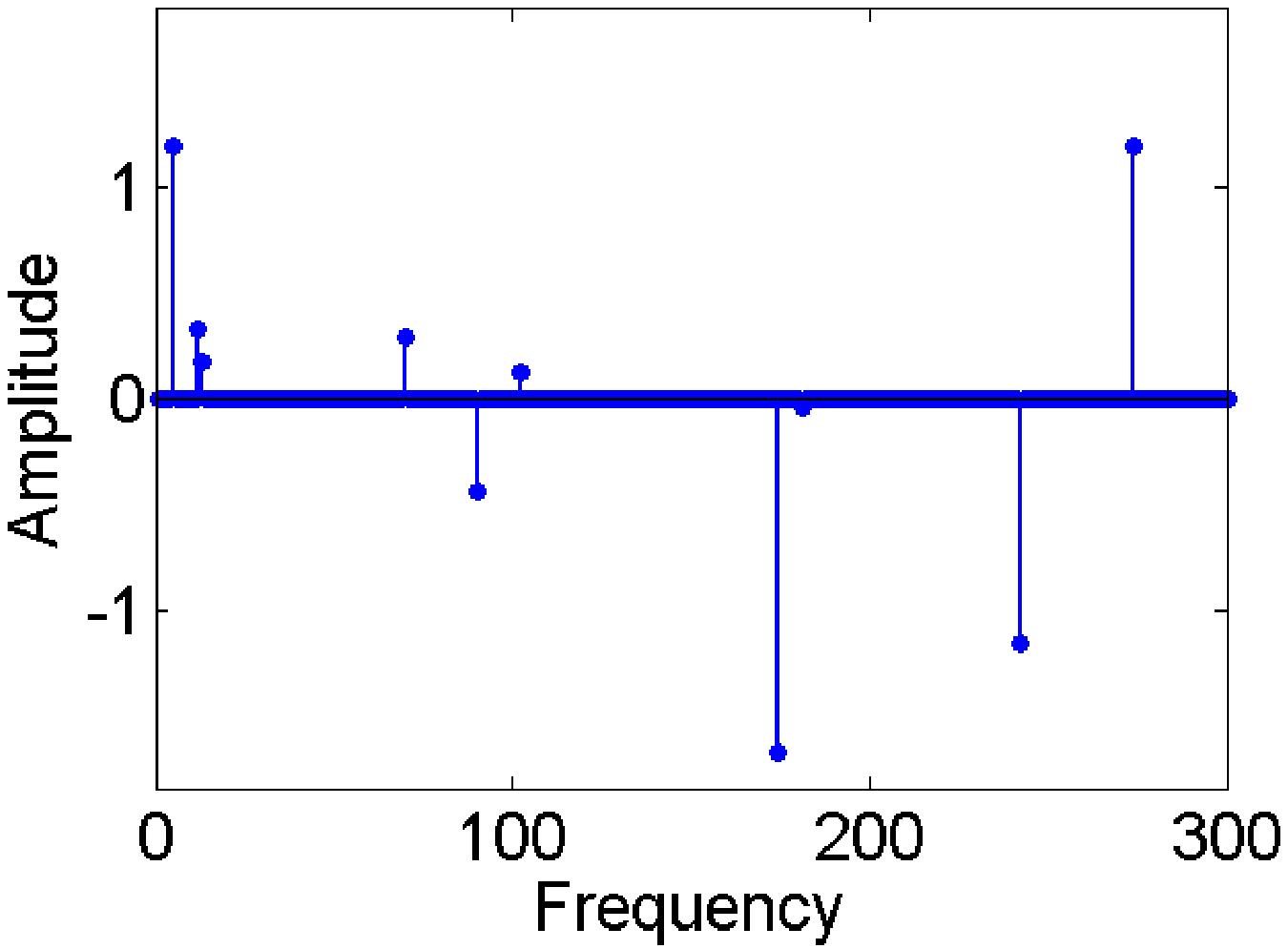}

\begin{center} Fourier-Coefficients \end{center}  
}
\parbox[t]{0.48\textwidth}{
\includegraphics[width=0.48\textwidth]{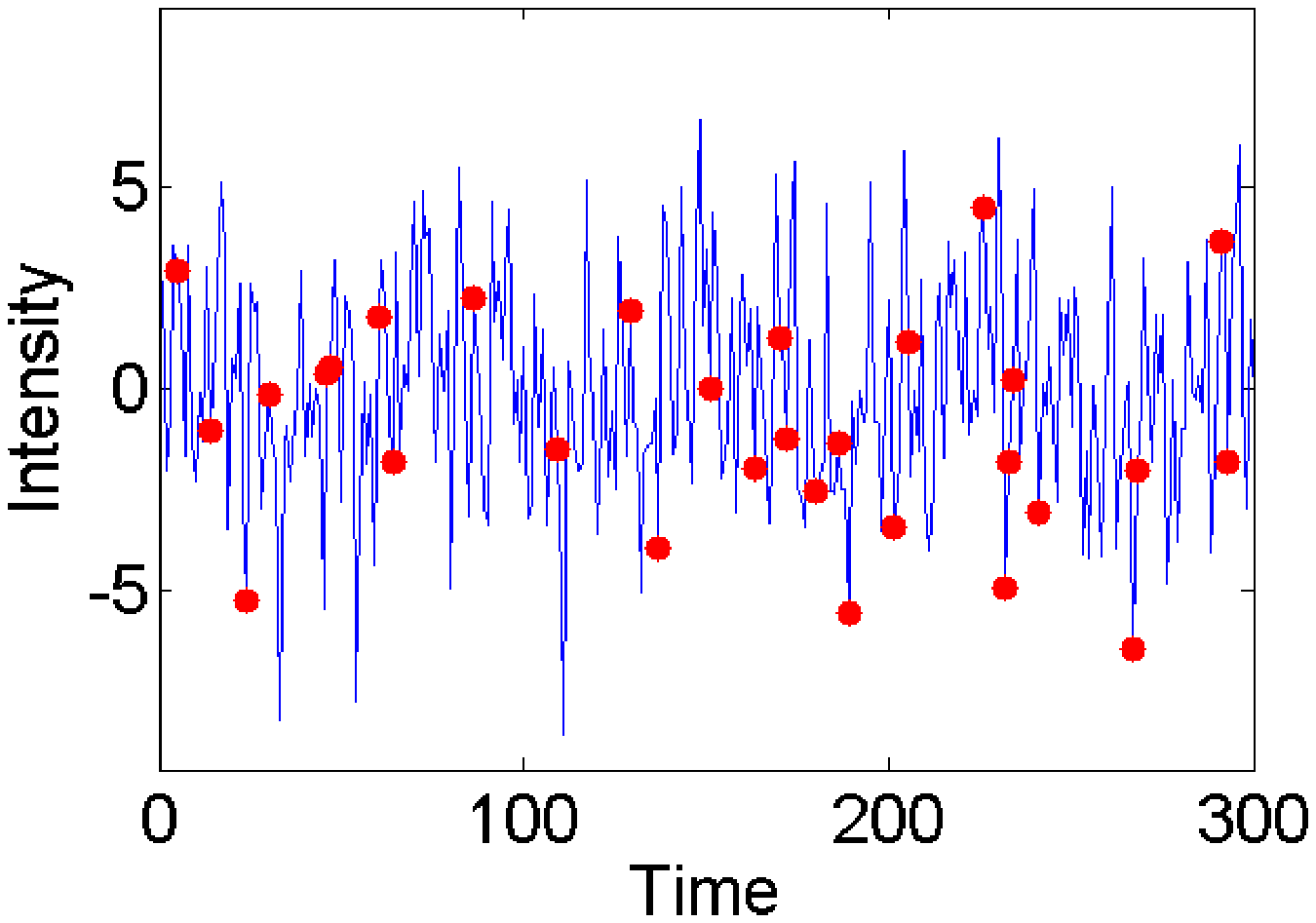}

\begin{center} Time-domain signal with 30 random samples \end{center}  
}\\
\parbox[t]{0.48\textwidth}{
\includegraphics[width=0.48\textwidth]{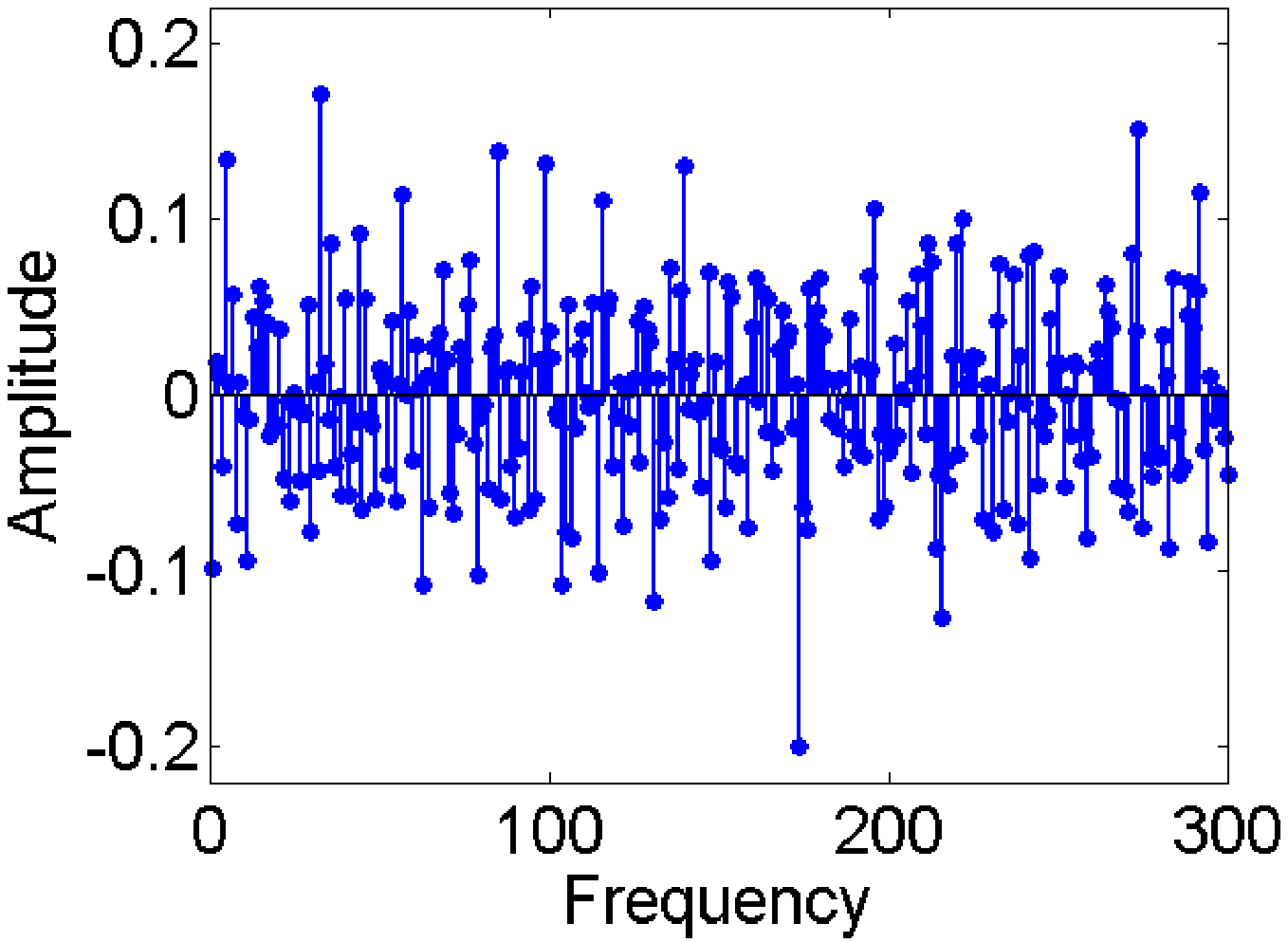}

\begin{center} Traditional least squares reconstruction  
\end{center}
}
\parbox[t]{0.48\textwidth}{
\includegraphics[width=0.48\textwidth]{BPSpectrum.eps}

\begin{center} Compressed sensing reconstruction \end{center}
}
\end{center}
\caption{Sparse recovery of sparse Fourier expansion}
\end{figure}

\begin{figure}
\begin{center}
\parbox[t]{0.3\textwidth}{
\includegraphics[width=0.3\textwidth, height=3.2cm]{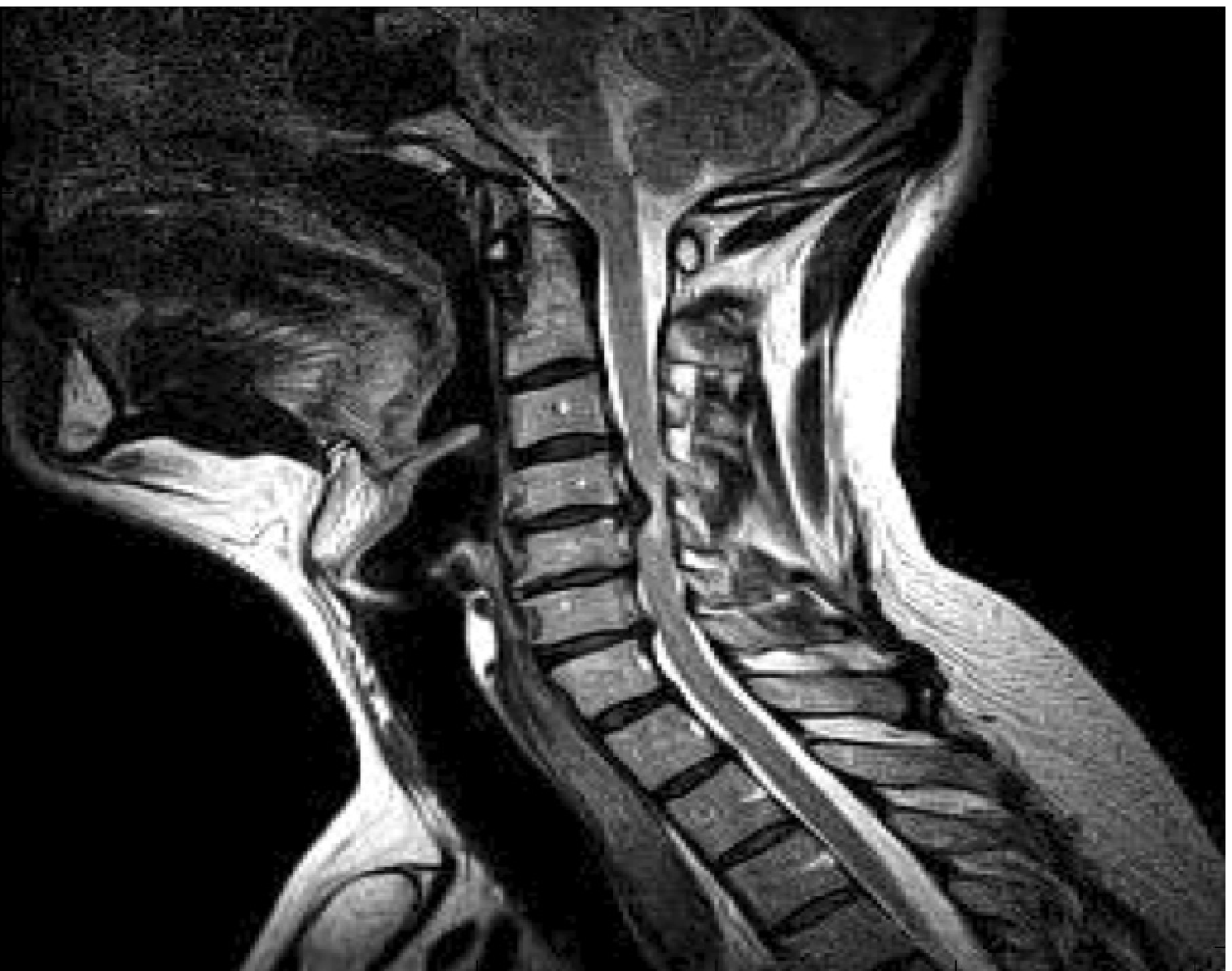}

\begin{center} Original image \end{center}  
}
\parbox[t]{0.3\textwidth}{
\includegraphics[width=0.3\textwidth, height=3.2cm]{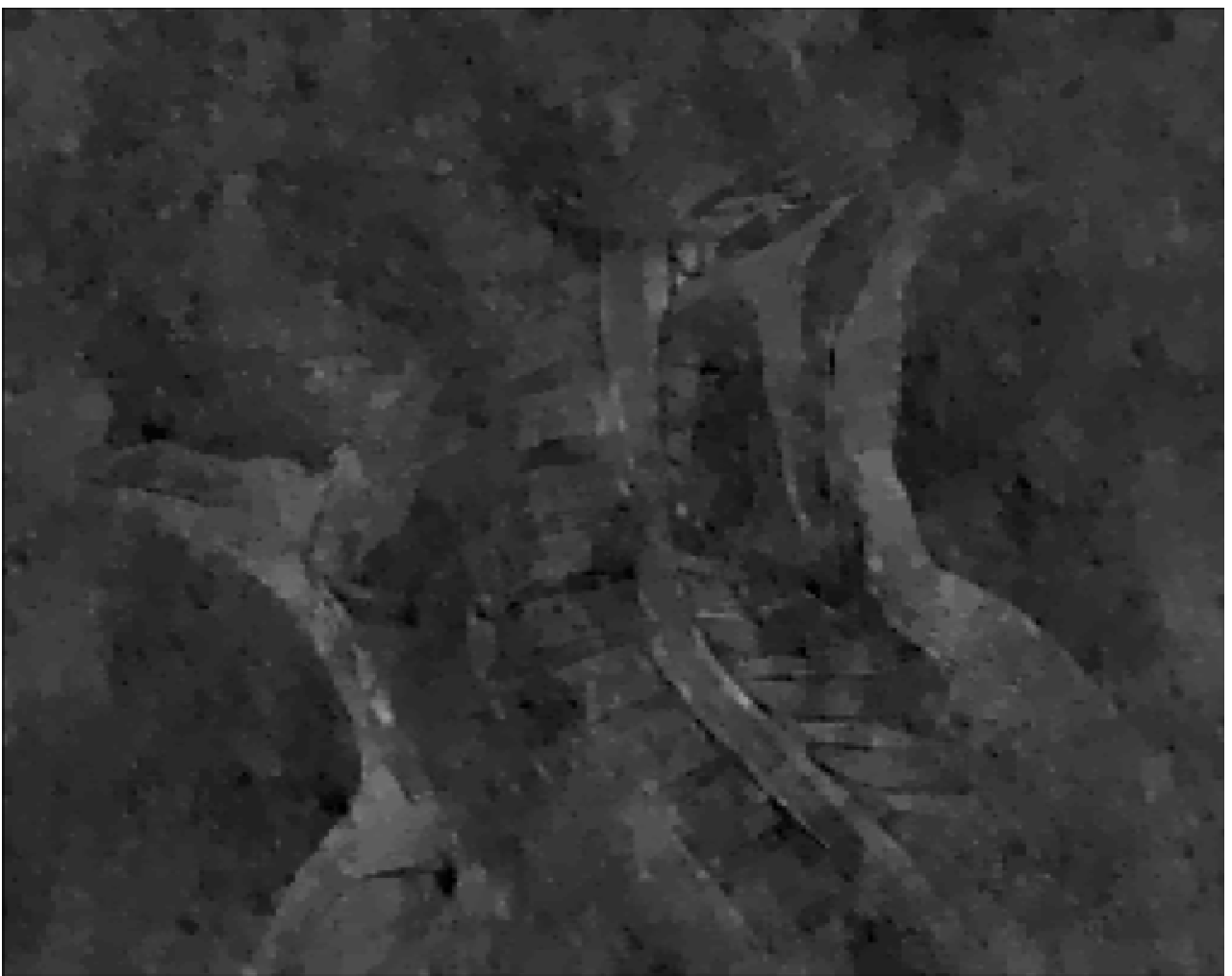}

\begin{center} Recovery from uniform Fourier samples \end{center}
}
\parbox[t]{0.3\textwidth}{
\includegraphics[width=0.3\textwidth, height=3.2cm]{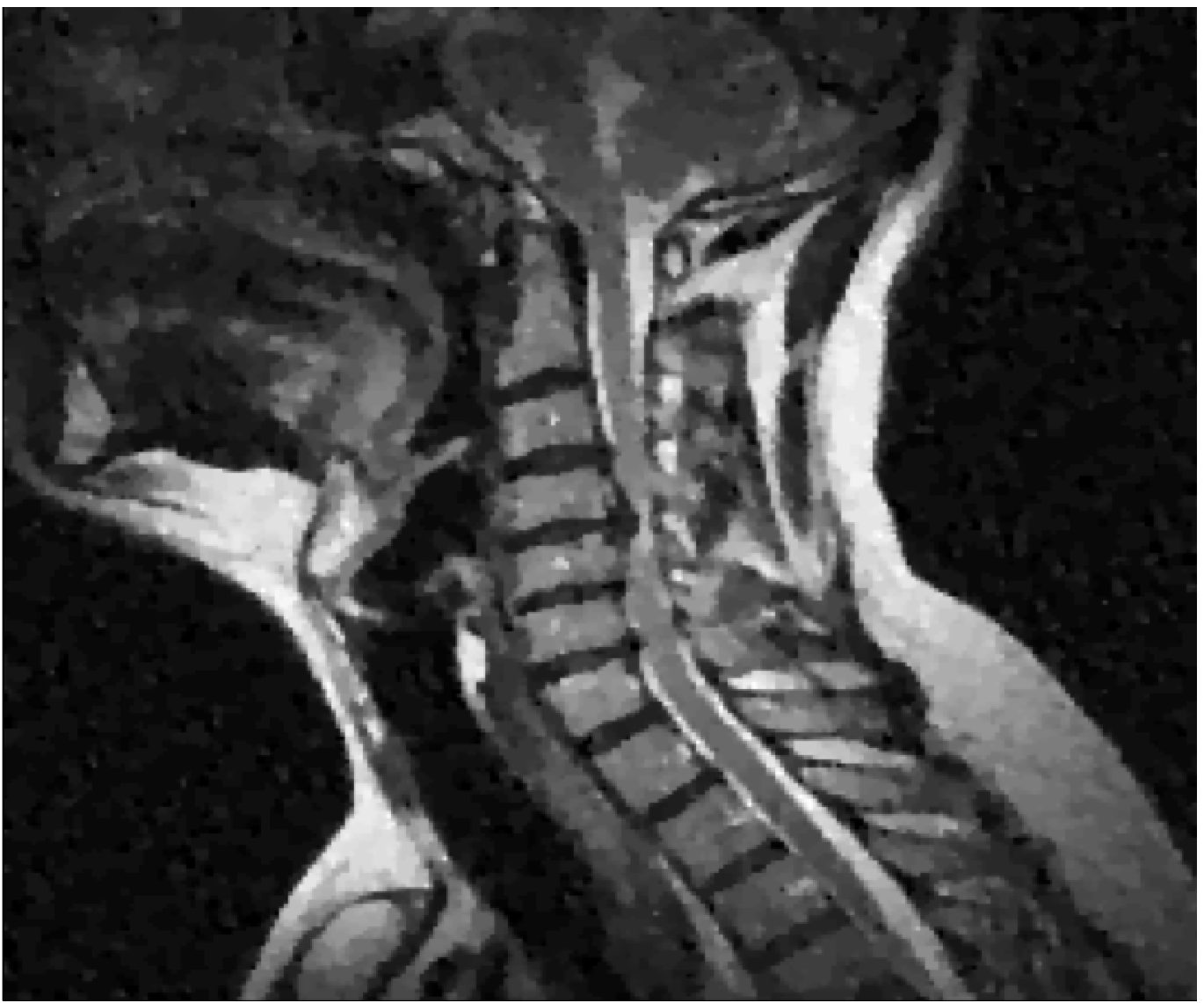}

\begin{center} Recovery from variable density Fourier samples \end{center}
}
\end{center}
\caption{Recovering a $256\times 256$ MRI image from $6400$ Fourier samples using Total Variation minimization \cite{krwaXX}. We compare a uniform sampling distribution to a sampling density proportional to $(\ell_1^2+\ell_2^2)^{-1}$.}\label{fig:spine}
\end{figure}

Figure~\ref{fig:SparseFourier} illustrates the recovery of a sparse Fourier expansion from few random samples via compressive sensing techniques, and for comparison also
shows a traditional reconstruction technique which clearly performs very poorly. Figure~\ref{fig:spine} considers the practical example of a $256\times 256$ MRI spine image, which is reconstructed from $6400$ Fourier samples, that is, less than $10\%$ of the information. It shows the need for variable density sampling schemes as they form the basis of Theorem~\ref{thm2} below. 

For {\em low-rank matrix recovery}, one also considers linear measurements, but the structural signal model is that the signal is approximated by a low-rank matrix. 
This problem closely relates to applications in recommender systems and signal processing \cite{ahro13,ahrero12,care09,coboma11}, but also has connections to quantum physics \cite{beeiflgrli09,gr09-2}.

In {\em phaseless estimation}, only the modulus of each measurement is observed. Such a measurement setup can be found in various applications in physics, 
such as diffraction imaging and X-ray crystallography. As the phases of an image are known to carry most of its information, this is a difficult problem, even when a number of measurements considerably larger than the dimension is used. Besides the general setup with no structural assumptions on the signal, the case of sparse signals has also in the phaseless estimation problem received considerable attention.

A major reason for the large research activity in compressive sensing and its extensions in the recent years is due to its potential for a large number of applications in signal processing and beyond.
Besides MRI (as illustrated in Figure~\ref{fig:spine} and described in more detail below), compressive sensing has applications to various signal processing applications. Let us mention a few.
Compressive sensing may improve several types of radar imaging. Especially when observing the sky, 
the fact that usually only a very limited number of airplanes is present at a time, naturally leads to sparsity of the image to be reconstructed. 
Moreover, the radar measurements can often be designed such that good measurement matrices
for compressing arise. This has been worked out for delay/Doppler radar \cite{hest09,pfra10,pfratr11}, for a setup with multiple antennas at random locations \cite{fastya09,hurast12}, 
for sparse MIMO radar \cite{frst12}, and more \cite{en10-1}. Another promising signal processing application of compressive sensing appears in microscopy for materials science applications \cite{cahusukeba11}. 
It is also possible to implement certain measurement matrices directly on CMOS chips, 
so that cameras may operate with fewer measurements or may increase resolution by keeping the costs down \cite{java09,hoxitaba13,rogrchroha10}. 
This may also be useful in order to reduce the power consumption of imaging sensors. Further optical and imaging applications of compressive sensing are described for instance in the overview 
articles \cite{maniwi11,ro08-1}. Further applications of compressive sensing and its extensions will be described in the individual sections below.

In the remainder of this paper, we will consider each of these scenarios separately, giving an overview over recovery algorithms and reconstruction guarantees for structured measurements. 
In Section~\ref{sec:CS}, we discuss compressed sensing, in Section~\ref{sec:MR}, we consider low rank matrix recovery, and the topic of Section~\ref{sec:PE} is phaseless estimation. In Section~\ref{sec:MM}, 
we give a brief overview of the mathematical proof techniques used to achieve the results, which are somewhat similar in all the three areas. We concentrate on structured random measurements in this article 
rather than giving a basic general introduction to the field. For introductory survey
papers on compressed sensing we refer to \cite{ba07-2,cawa08,fora11,ra10} and for recent books to \cite{elku12,fora13}.

\section{Compressed Sensing} \label{sec:CS}

Throughout most of this section, we focus on the following setup. Note that we consider complex-valued signals and matrices, but the reader may as well 
imagine just the real case. 
\begin{itemize}
 \item {\bf Signals:} Let $x\in\K^N$ be approximately $k$-sparse in a basis ${\mathcal B}=\{b_i\}_{j=1}^N$, that is, 
 $x \approx \sum_{j \in S} x_j b_j$, for $S \subset [N]:= \{1,2,\hdots,N\}$ of cardinality $k$.
 The important case that the representation system is a frame rather than a basis will not be covered in this survey, we refer the reader 
 to \cite{rascva08}, \cite{caelnera11}, and many follow-up papers. To quantify the sparse approximation quality, we define for $0<p<\infty$ the $\ell_p$ best $k$-term 
 approximation error $\sigma_k(x)_p:= \inf\limits_{\substack{z\in\K^N : z\text{ is } k-\text{sparse}}} \|x-z\|_{p}$. A common model for approximately sparse vectors is given 
 by the unit $\ell_p$ ball, $p<1$. Namely, $\|x\|_{\ell_p}\leq 1$, $p<1$, is known to imply $\sigma_k(x)_1\leq k^{1-1/p} \|x\|_{p}$ \cite{fora13}.  As $p\rightarrow 0$, these quasinorms converge 
 to the support size $\|x\|_0:=|\{j:x_j\neq 0\}|$. For all of these concepts, when no sparsity basis is specified or clear from the context, we will work with the standard basis. 
 Note that in contrast  to a noisy sparse signal, there is no underlying sparse ``truth''. The goal will be to estimate the approximately sparse signal and the result will not necessarily have to be sparse either.
 \item {\bf Measurements:} We consider $m$ linear measurements, represented in matrix form by $A\in \K^{m\times N}$. 
The measurements are affected by additive noise $e\in\K^N$. Thus the observed measurements are given by $y=Ax+e \in\K^m$. 
We will mainly consider adversarial noise, i.e., we are looking for worst case bounds for the reconstruction error. 
However, random noise models have also been considered, mainly in a statistical context, see e.g.\ \cite{birits09,cata07}.
\end{itemize}
As mentioned in the introduction, the general paradigm that we will follow in all three application scenarios will be to design the measurements in a random fashion such that efficient reconstruction can be guaranteed. As such, the resulting reconstruction guarantees will be probabilistic, namely reconstruction is only guaranteed with high probability. There are two fundamentally different interpretations of such probabilistic guarantees. {\em Uniform recovery guarantees} concern all (approximately) sparse signals at the same time. That is, one seeks random matrix construction whose realizations, with high probability, allow for the recovery of all sparse signals. {Non-uniform recovery guarantees}, on the other hand, establish that for any given signal, recovery is possible with high probability. That is, in the latter case, the matrices for which reconstruction fails can differ for different signals and there is no guarantee that one matrix can be generated that allows for the recovery of all sparse vectors.

The techniques used to establish uniform versus non-uniform recovery guarantees are somewhat different. 
An important tool that has been successfully used to establish uniform recovery guarantees is the restricted isometry property.
\begin{definition}
 A matrix $A\in\K^{m\times N}$ is said to have the restricted isometry property of order $k$ and level $\delta$ with respect to a basis $\mathcal B$ (in short, the $(k,\delta)$-RIP), if it satisfies
 \begin{equation}
  (1-\delta)\|x\|^2_2 \leq \|Ax\|^2_2 \leq (1+\delta)\|x\|^2_2 \label{eq:RIP}
 \end{equation}
for all $x\in \K^N$ which are $k$-sparse with respect to $\mathcal B$. The smallest $\delta$ that satisfies \eqref{eq:RIP} is called the restricted isometry constant of order $k$ and denoted by $\delta_k$.
\end{definition}
If the measurement matrix has restricted isometry property of order $k$ and suitably small level $\delta$, then reconstruction of sparse vectors can be guaranteed for various algorithms. 
A simple tractable recovery approach which is arguably best understood in the context of compressed sensing is $\ell_1$-minimization (basis pursuit) \cite{chdosa99}. 
Here it is assumed that the noise level $\epsilon=\|e\|_2$ is known at least approximately. Furthermore, $B$ denotes the change of basis matrix associated to the sparsity basis $\mathcal B$. Then the resulting minimization problem is as follows,
\begin{equation}
 \widehat{x}= \argmin\limits_{x:\|Ax-y\|_2\leq \epsilon} \|Bx\|_{1} \tag{$\ell_1$}. \label{eq:l1}
\end{equation}
It may not be obvious that $\ell_1$-minimization promotes sparsity, but there are many theoretical results indicating this fact in general, see e.g.\ Theorem 3.1 in \cite{fora13}. 
More specifically, the following theorem proved in \cite{cazh14} shows that this convex optimization problem successfully recovers approximately sparse solutions provided the
measurement matrix has a sufficiently small restricted isometry constant.
\begin{theorem}\label{thm:RIPrc}
 Let $x\in\K^N$, assume $A\in\K^{m\times N}$ has the restricted isometry property of order $2k$ and level $\delta < \tfrac{1}{\sqrt{2}}\approx 0.707$ with respect to the basis $\mathcal B$, and let $y=Ax+e$, where the noise vector $e$ satisfies $\|e\|_2\leq \epsilon$. Then the vector $\widehat x$ recovered by the minimization problem \eqref{eq:l1} satisfies
 \begin{equation}
  \|\widehat{x}-x\|_2\leq C_1 \frac{\sigma_k(Bx)_1}{\sqrt{k}} + C_2\epsilon. \label{eq:RIPrec}
 \end{equation}
Here $C_1$ and $C_2$ are absolute constants.
\end{theorem}
Note that for $\epsilon=0$, the theorem implies that under the same conditions on $A$, 
every $k$-sparse vector is recovered exactly by \eqref{eq:l1} provided the measurements are not affected by noise. 
The constant $1/\sqrt{2}$ above is optimal \cite{dagr09} and the error bound \eqref{eq:RIPrec} as well, see also below.
Moreover, also error bounds in $\ell_p$ with $1\leq p \leq 2$ can be shown \cite{fora13}.


Similar recovery guarantees, though for smaller thresholds for $\delta$, have been derived for other recovery algorithms than $\ell_1$-minimization.
Examples include CoSaMP, Iterative Hard Thresholding and Iterative Hard Thresholding Pursuit \cite{fora13}. 

{\bf Subgaussian random matrices.} In order to deduce recovery results using these results, the measurement systems under consideration must have restricted isometry constants below some constant threshold. This can be achieved by choosing a measurement matrix with independent entries drawn according to a subgaussian distribution as given in the following definition (see for example \cite{ve12} for a detailed discussion of subgaussian random variables including a number of equivalent definitions).

\begin{definition}
 A real or complex random variable $X$ is called subgaussian with parameter $\beta > 0$ if 
 \[
 \P(|X| \geq t) \leq 2 e^{-\beta t^2}.
 \]
 A random matrix $A$ is called subgaussian with parameter $\beta$ if its entries are independent mean zero subgaussian random variables with parameter $\beta$.
\end{definition}
Examples of subgaussian random variables include centered Gaussian random variables and centered bounded random variables such as Rademacher random variables, i.e., $\P(\xi=\pm 1)=\tfrac{1}{2}$.

The following theorem concerning the RIP for subgaussian random matrices is well-known, 
a particularly simple proof can be found in \cite{badadewa08}, see also
\cite{fora13}. This and most following results in this survey are probabilistic, that is, they hold with high probability on the draw of the measurement matrices. 
The precise meaning of this is that we can make the probability of failure arbitrarily small by (slightly) increasing the number $m$ of samples.
For reasons of simpler presentation, we will not specify this dependence, but refer to the original research articles.
\begin{theorem}
 Let $\mathcal B$ be an orthonormal basis of $\,\K^N$, $A\in\K^{m\times N}$ be a subgaussian random matrix with parameter $\beta$, and assume that 
 $m\geq C_\beta \delta^{-2} (k\log\big(\tfrac{n}{k}\big) + \log(\gamma^{-1}))$. Then with probability at least $1-\gamma$,  the matrix $\tfrac{1}{\sqrt{m}} A$ 
 has the restricted isometry property of order $k$ and level $\delta$ with respect to the basis $\mathcal B$. Here 
 $C_\beta$ is a constant, which only depends on $\beta$.
\end{theorem}

Combined with Theorem~\ref{thm:RIPrc}, this result yields recovery guarantees for $\ell_1$-minimization for embedding dimensions $m\geq C_\beta'  k\log\big(\tfrac{n}{k}\big)$. 
In particular, if $k$ is much smaller than $n$ then also the number $m$ of measurements can be chosen smaller than the signal length $n$ and still signal recovery is ensured.
An embedding dimension of order $k\log\big(\tfrac{n}{k}\big)$ is known to be necessary to achieve recovery guarantees of the form \eqref{eq:RIPrec} \cite{codefora11}. 
For the case of subgaussian matrices, also the non-uniform approach mentioned above does not yield recovery guarantees for embedding dimensions of a smaller order. 
Note that the result is universal in the sense that the recovery guarantees do not depend on the choice of $\mathcal B$.

Subgaussian compressed sensing matrices are often considered a benchmark to judge the quality of a randomized construction. Thus one is typically interested in embedding dimensions which scale linearly in the sparsity $k$, up to logarithmic factors. Deterministic constructions to date are nowhere near this benchmark. Efficient constructions are known for embedding dimensions which scale quadratically in $k$ \cite{fora13}, the best known infinite family that beats this quadratic bottle neck uses heavy machinery from additive combinatorics to achieve a scaling of $k^{2-\mu}$ \cite{bodifokoku11}, where the best currently available estimate of $\mu$ is on the order of $10^{-26}$ \cite{MixonBlog}.

{\bf Random Fourier measurements.} From the very beginning, one of the main motivating applications of compressed sensing was magnetic resonance imaging (MRI). MRI measurements are known to be well modeled with the (continuous) Fourier transform. We will follow the common approach to approximate this setup by discrete Fourier transform (DFT) measurements. In this article we work with the non-normalized discrete Fourier transform matrix with entries given by $F_{j\ell}=e^{-2\pi i \frac{j \ell}{N}}, \ -\tfrac{N}{2}+1 \leq j, \ell \leq \tfrac{N}{2}$ (where we assume $N$ to be even).
The discrete Fourier basis then consists of the normalized rows of $F$. This discretization approach also has the advantage that the resulting (approximate) measurement matrix can be efficiently computed using the Fast Fourier Transform (FFT). We note, however, that refined models for compressed sensing MRI have been developed; we mention infinite dimensional compressed sensing \cite{adha11} and discrete prolate spheroidal sequences \cite{dawa12}. 

In the undersampled setup that we consider here, only a subset of the rows of the DFT are chosen. Randomness is introduced into the model by selecting these rows at random. Note that this simple subsampling model does not incorporate certain technical side constraints of the MRI acquisition process. For example, samples are in practice acquired on continuous trajectories, which is not in line with drawing the samples at random. Building this and other technical constraints into the compressed sensing model remains an active area of research, see for example \cite{chcikawe14} for a recent attempt to address the continuity constraint. In this survey, we will, however, stick with the simplified model of randomly chosen DFT measurements.

Because of the structure of the resulting random matrix, the recovery guarantees must depend on the basis in which the signal is sparse. This is most easily seen by considering signals sparse in the discrete Fourier basis. Then most DFT coefficients of a sparse signal are zero by construction, so one requires a number of measurements much larger than what is needed for subgaussian measurement matrices. As it turns out, a sufficient property to ensure recovery guarantees is {\em incoherence} between the sparsity basis and the measurement basis (in this case, the discrete Fourier basis).
\begin{definition}
 The {\em coherence} between two orthonormal bases ${\mathcal B}_1$ and ${\mathcal B}_2$ of $\K^N$ is defined as
 \begin{equation}
  \mu({\mathcal B}_1,{\mathcal B}_2) =\sup_{b_1\in{\mathcal B}_1,b_2\in{\mathcal B}_2} |\langle b_1, b_2\rangle|.
 \end{equation}
\end{definition}
A main example for incoherent bases are the standard basis and the Fourier basis, their coherence has the minimal value of $N^{-1/2}$.
Though often formulated specifically for the standard basis as the sparsity basis and the Fourier basis as the measurement bases, the uniform recovery results directly generalize to arbitrary incoherent bases. The first results of the type below were introduced in \cite{cata06,carota06-1} and later refined in \cite{ruve08,ra08,chguve12,rawa13}. 
\begin{theorem}
\label{thm:incoh} 
Consider orthonormal bases ${\mathcal B}_1$ and ${\mathcal B}_2$ of $\K^N$ with coherence bounded by
$\mu({\mathcal B}_1,{\mathcal B}_2) \leq \KK N^{-1/2}.$
Fix \HR{$\delta, \gamma  \in (0,1)$} and integers $N, m$, and $k$ such that
\begin{equation}\label{incoh:RIP:cond}
\HR{m \geq C \delta^{-2} \KK^2 k \max\{\log^3(k) \log(N), \log(\gamma^{-1}) \}.}
\end{equation}
Consider the matrix $\Phi \in \K^{m \times N}$ formed by subsampling $m$ vectors of ${\mathcal B}_2$ independently according to the uniform distribution. 
\HR{Then $A=\frac{\sqrt{N}}{\sqrt{m}}\Phi$ has the restricted isometry property of order $k$ and level $\delta$ with respect to the sparsity basis ${\mathcal B}_1$ with probability at least $1-\gamma$}. 
The constant $C>0$ is universal (independent of all other quantities).
\end{theorem}

Note that the normalization factor implies that the rows rather than the columns are (approximately) normalized. This renormalization is necessary as the restricted isometry property requires that the columns are approximately unit norm.

In contrast to the case of subgaussian matrices, the best known recovery guarantees for the non-uniform approach are considerably stronger than in the uniform case. Namely, the number of measurements required to guarantee recovery of $k$-sparse signals in $N$ dimensions with high probability is of order $k \log(N)$ (see for example \cite{carota06, ra07, ra10, capl11-1,fora13}).

The direct applicability of Theorem~\ref{thm:incoh} is somewhat limited. While for certain applications, such as angiography, sparsity in the standard basis can be assumed, most sparsity inducing representation systems for images, such as wavelets or shearlets are not incoherent with the Fourier basis. For Haar wavelets, for example, the constant Fourier mode and the constant wavelet mode even agree, so the bases are maximally coherent, that is, $\mu=1$ or, in the notation of Theorem~\ref{thm:incoh}, $\KK=\sqrt{N}$. In contrast to the case of sparsity in the Fourier basis, where hardly any subsampling is possible, this high coherence only concerns very few of the measurement vectors. Most measurement vectors have uniformly small inner products  with all vectors in the sparsity basis. This can be exploited by sampling according to a variable density based on a localized refinement of the coherence.  In this way, one can obtain recovery guarantees for Fourier measurements and images approximately sparse in the 
Haar wavelet basis. As the result for noisy measurements becomes quite technical, 
we only state the recovery guarantees without noise and refer to \cite{krwaXX} for details on the noisy case. In the following result, $\mathcal H$ denotes the Haar wavelet transform matrix. Due
to its appearance we restrict $N$ to be a power of $2$ for notational simplicity.
\begin{theorem}[\cite{krwaXX}]
\label{thm2}
Fix integers \HR{$N = 2^p, m, k$ and $\gamma \in (0,1)$} such that 
\begin{equation}
\label{m:wav}
\HR{m \geq C_1 k \max\{\log^{3}(k)\log^2(N), \log(\gamma^{-1})\}}.
\end{equation}
Select $m$ frequencies $\Omega = \{ (\omega_1^j, \omega_2^j ) \}_{j=1}^m \subset \{-N/2+1, \dots, N/2\}^2$ independently according to
\begin{align}
\label{inverseD}
\P &\big[ (\omega_1^j, \omega_2^j) = (\ell_1, \ell_2) \big]  = C_N \min\left(C, \tfrac{1}{\ell_1^2 + \ell_2^2 }\right),\ -\tfrac{N}{2}+1 \leq \ell_1, \ell_2 \leq \tfrac{N}{2},
\end{align}
where $C$ is an absolute constant and $C_N$ is chosen such that \eqref{inverseD} defines a probability distribution.  

Then \HR{with probability at least $1-\gamma$}, the following holds for all images $f \in \C^{N \times N}$:  Given measurements $y = {\mathcal F}_{\Omega}f$, the estimation
\begin{equation}
\label{l1-Haar}
f^{\#} = \argmin_{g \in \C^{N \times N}} \| {\mathcal H}g \|_1  \quad  \textrm{such that}  \quad  {\mathcal F}_{\Omega} g = y
\end{equation}
approximates $f$ up to the best $k$-term approximation error in the bivariate Haar basis: 
\begin{equation}
\label{stable3-Haar}
\| f -f^{\#} \|_2 \leq C_2 \frac{ \sigma_k({\mathcal H}f)_1}{\sqrt{k}}.
\end{equation}
\end{theorem}
\FK{It should be noted that while it is not clear whether the decay of second argument of the $\min$ in \eqref{inverseD} can be improved (maybe for different families of wavelets), the cut-off introduced by the first argument is crucially tied to our sampling model. As we are sampling with replacement, it is necessary to prevent frequencies near the origin from being sampled too often.}

Similar recovery guarantees can be derived for images with approximately sparse discrete gradients, where the recovery algorithm is based on total variation minimization (see \cite{krwaXX} for details). Figure~\ref{fig:spine} above illustrates the need for variable density sampling by comparing a uniform sampling density and an inverse square density.  

In parallel to this refinement of the concept of Fourier/wavelet incoherence, the concept of sparsity has also been refined to better reflect the fact that wavelet coefficients on larger scales exhibit less sparsity than on smaller scales. In \cite{rawa13}, the authors study a weighted sparsity model and derive RIP-based guarantees for uniform recovery. In \cite{adhaporo13}, the authors work in an infinite-dimensional setup and consider both coherence and sparsity in the asymptotic limit. These observations may also serve as an explanation that empirically, sampling densities with a faster decay seem to outperform those predicted by Theorem~\ref{thm2}.

{\bf Subsampled random convolutions.} Later, further applications of compressed sensing arose that require different structural constraints on the measurements. For application in remote sensing \cite{hest09,ro09-1} and coded aperture imaging \cite{mawi08}, the model of choice is often that of subsampled convolutions. For simplicity, we consider the circular convolution $(x,\xi)\in\K^N\times\K^N \mapsto x*\xi\in \K^N$ given by
\begin{equation}
 (x*\xi)_j = \sum_{i=1}^N x_i \xi_{j\ominus i},
\end{equation}
where $\ominus$ denotes subtraction mod $N$. These convolution measurements are then subsampled by an operator $P_\Omega$, $\Omega\subset \{1, \dots N\}$, which restricts a vector to only those entries indexed by $\Omega$. That is, after normalizing the columns, we obtain a measurement matrix $A\in\K^{m\times N}$ of the form
\begin{equation}
 Ax:=\tfrac{1}{\sqrt{m}} P_\Omega (\xi*x). \label{eq:partcirc}
\end{equation}
A matrix $A$ of this form is called a {\em partial circulant} matrix.
For this setup there are two alternative ways to introduce randomness into the system. One can again choose the subsampling pattern $\Omega$ at random, or one can choose $\Omega$ deterministically and randomize the vector $\xi$. For both scenarios, recovery guarantees have been derived. 
In the case of random $\Omega$, the key concept determining the success is the {\em autocorrelation} of $\xi$. Namely, as shown in \cite{ligali13}, a sufficient condition is that $\xi$ is a {\em nearly perfect sequence} in the sense that  for $\ell\in \{1,\dots N-1\}$ its off-peak autocorrelation, that is, the values
\begin{equation}
 R_\xi(\ell) = \sum_{j=1}^{N} \xi_j \bar{\xi}_{j\oplus \ell}
\end{equation}
for $\ell$ not a multiple of $N$,
is bounded  by a small constant $c$ independent of $N$. Here $\oplus$ denotes addition mod $N$. With this definition, the result in \cite{ligali13} reads as follows.
\begin{theorem}
 Let $\xi\in \K^N$ be a nearly perfect sequence, $|R_\xi(\ell)| \leq c$ for all $\ell$ not a multiple of $N$, such that $|(F\xi)_i|=\sqrt{N}$ for $i\in\{1,\dots, N\}$ and assume \HR{$m\geq C (1+c) \delta^{-2} k \max\{\log^4(N), \log(\gamma^{-1})\}$}, where $C$ is an absolute constant. Then with \HR{probability at least $1-\gamma$} 
 the matrix $A\in\K^{m\times N}$ as defined in \eqref{eq:partcirc} has the restricted isometry property of order $k$ and level $\delta$ both with respect to the standard and the Fourier basis. 
\end{theorem}
The proof is closely related to the case of random partial Fourier matrices as discussed above. In a similar way, one also obtains non-uniform recovery guarantees for embedding dimensions of order $k \log(N)$. For a detailed discussion of examples for nearly perfect sequences, we refer the reader to \cite{ligali13} and the references therein.

As in the partial Fourier case, this setup is based on random subsampling. That is, the sampling setup cannot be truly designed for lower sampling frequencies, as with very high probability, {\em some} subsequent samples will be selected. For this reason, the setup of choosing $\Omega$ deterministically and randomizing $y$ has received considerable attention. In coded aperture imaging, this corresponds to choosing a random aperture pattern, and in remote sensing, this concerns a random pulse that is transmitted. Both are often considerably easier to implement than subsampling at random.

The first results on this problem required embedding dimensions significantly worse than the linear benchmark scaling in $k$, namely cubic \cite{bahanorawr07}, quadratic \cite{bahanora10}, and $k^{3/2}$ \cite{rarotr11}. In \cite{krmera12}, linear scaling up to logarithmic factors has been achieved, as given in the following result. Here the random vector $\xi$ consists of independent subgaussian entries. Again, this includes the important examples of a Gaussian and a Rademacher random vector.
\begin{theorem}\label{thm:KMR}
 Let $\xi \in \K^N$ be a random vector with independent subgaussian entries with parameter $\beta$ and let $A \in \K^{m \times n}$ be a draw of the associated partial random circulant matrix generated by $\xi$ as given in \eqref{eq:partcirc}. If
\begin{equation}\label{m:circRIPcond}
\HR{m \geq C_\beta \delta^{-2} k \max\{ (\log^2 k) (\log^2 n), \log(\gamma^{-1}\}},
\end{equation}
then $A$ has the RIP of order $k$ and level $\delta$ \HR{with probability at least $1-\gamma$}. The constant $C_\beta>0$ only depends on the subgaussian parameter $\beta$.
\end{theorem}
There are also non-uniform recovery guarantees available for subsampled convolutions \cite{ra09}. The best known such result again achieves a scaling of order $k \log(N)$ for the embedding dimension \cite{ra09,ra10,ja13}.

There have been various works on other structured random measurement matrices for compressed sensing. Time-frequency structured random matrices, a class of examples which arises in wireless communication and radar, have been studied in \cite{pfratr11,krmera12} (uniform recovery guarantees) and \cite{pfra10} (non-uniform recovery guarantees). For a class of matrices which arises in radar imaging with randomly located antennas, only non-uniform guarantees are available to date \cite{hurast12}.

\section{Low Rank Matrix Recovery} \label{sec:MR}

Low rank matrix recovery represents an interesting extension of compressive sensing, where the sparsity assumption is replaced by a low rank assumption.
More precisely, the task is to reconstruct a matrix of low rank (or approximately low rank) from incomplete linear measurements. This problem arises for instance in recommender system design \cite{care09} 
\HR{and a number of  signal processing applications, described briefly next. 

Suppose that several sensors observe different aspects of the same phenomenon, for instance weather time series at locations distributed in some area.
Collecting the signals corresponding to each sensor data as columns into a data matrix, correlatedness of the data may lead to an (approximately) low rank of this matrix.
Therefore, low rank matrix recovery techniques help in such situations to accurately reconstruct signals from observed data and/or to work with fewer sensor measurements, see  \cite{ahro13} .  
In wireless communications it is a main task to estimate both the transmission channel as well as the sent signal from the received signal. In \cite{ahrero12}, this blind deconvolution 
problem is reformulated as a low rank matrix recovery problem and recovery guarantees for this scenario are shown. 
Another important task arising in so-called cognitive radio is to decide whether certain frequency bands are occupied or not so that free bands may be potentially used
for wireless transmission by a user. In \cite{coboma11} an approach to this problem via low rank matrix recovery was introduced.}

In mathematical terms, the measurements of a matrix $X \in \C^{n_1 \times n_2}$ are provided by a linear map ${\mathcal A} : \C^{n_1 \times n_2} \to \C^m$, i.e.,
\[
y = {\mathcal A}(X).
\]
We are interested in the underdetermined case that $m < n_1 n_2$. A prominent special case is the matrix completion problem, where one samples entries of a low rank
matrix and tries to fill in the missing entries, see also below.

The rank minimization problem 
\[
\min \operatorname{rank}(X) \quad \mbox{ subject to } {\mathcal A}(X) = y
\]
is unfortunately NP-hard. (In fact, the NP-hard problem of finding for an underdetermined linear system the solution with the smallest support -- as it appears in compressed sensing context -- can be cast as a rank-minimization problem.) 
Let $\sigma(X)=(\sigma_1(X),\sigma_2(X),\hdots,\sigma_n(X))$, $n = \min\{n_1,n_2\}$, be the vector of singular values of $X$.
Observing that $\operatorname{rank}(X) = \|\sigma(X)\|_0$ and having the $\ell_1$-minimization approach for the standard compressed sensing problem in mind,
we are led to consider the nuclear norm
\[
\| X \|_* = \|\sigma(X)\|_1 = \sum_{j=1}^n \sigma_j(X),
\]
and the corresponding nuclear norm minimization problem \cite{fa02-2,fapare10}
\begin{equation}\label{nucl:min}
\min \|X\|_* \quad \mbox{ subject to } {\mathcal A}(X) = y. \tag{N}
\end{equation}
This is a convex optimization problem for which a number of algorithms exist \cite{cacash10,chpo11,goma11}. For instance, it can be reformulated as a semidefinite optimization program.
Other algorithmic approaches for the low rank matrix recovery problem including iterative hard thresholding \cite{ceky12,tawe13}, iteratively reweighted least squares \cite{forawa10}
and a variant of CoSaMP called ADMiRA can be pursued as well \cite{brle10}.

Similarly as in the standard compressed sensing case, we are interested in suitable measurement maps ${\mathcal A}$ and the minimal number $m$ of samples
required to reconstruct an $n_1 \times n_2$ matrix of rank $r$. It should not be a surprise by now that again, random measurement maps are optimal with high
probability in this context in the sense of working with a minimal
number  of measurements. We also distinguish uniform and nonuniform recovery guarantees here.

A prominent approach for deriving uniform guarantees consists in studying a version of the restricted isometry property for the low rank case \cite{fapare10}. 
Similarly as in \eqref{eq:RIP}
we define the restricted isometry constant $\delta_r$ to be the smallest number such that
\begin{equation}\label{rankRIP}
(1-\delta_r) \|X\|_F^2 \leq \| \mathcal{A}(X)\|_2^2 \leq (1+\delta_r) \|X\|_F^2 \quad \mbox{ for all } X \in \C^{n_1 \times n_2} \mbox{ with } \operatorname{rank}(X) \leq r.
\end{equation}
If $\delta_{2r} < 1/\sqrt{2}$, then nuclear norm minimization \eqref{nucl:min} uniquely recovers every matrix of rank at most $r$ from $y = {\mathcal A}(X)$. Moreover,
recovery is stable under noise on the measurements and passing to approximately low-rank matrices 
\cite{fapare10}.

The simplest model of a random measurement map is a Bernoulli or Gaussian map, 
where all the entries of the tensor ${\mathcal A}_{jk\ell}$, i.e., ${\mathcal A}(X)_j = \sum_{k,\ell} {\mathcal A}_{jk\ell} X_{k\ell}$, 
are chosen as independent mean-zero Rademacher or standard Gaussian random variables. The restricted isometry constant of the rescaled map $\tfrac{1}{\sqrt{m}} {\mathcal A}$
satisfies $\delta_r \leq \delta$ with probability at least \HR{$1-\gamma$} provided that \cite{capl11}
\[
\HR{m \geq C \delta^{-2} \left(r (n_1 + n_2)) + \log(\gamma^{-1}) \right)}.
\]
This implies rank-$r$ matrix recovery with high probability via nuclear norm minimization from $m \geq C r(n_1+n_2)$ measurements. This bound is optimal
as the right hand side represents essentially the number of degrees of freedom of a matrix of rank $r$ in dimension $n_1 \times n_2$. 
Clearly, if $r \ll \min\{n_1,n_2\}$, then $m$ can be chosen smaller than the dimension $n_1n_2$ of the underlying matrix space $\C^{n_1 \times n_2}$.

While Bernoulli and Gaussian measurement maps are comparably easy to analyze, they are of limited practical use due to lack of structure. 
As in the standard sparsity case, we therefore rather look for structured random measurement maps. 
(Optimal deterministic constructions are not available at this point and likely very hard to achieve.) 

\medskip

{\bf Matrix completion.} Imagine an online vendor system which asks clients to rate the products which they have purchased. Such recommendations can be organized in a big matrix
where the columns represent products and the rows the clients. The corresponding ratings are kept as the entries of this matrix. Since not every client rates every product, 
a lot of entries of the matrix are missing. For obvious reasons, a recommender system needs to guess such missing entries 
in order to make good suggestions of products that a client will probably like. In other words, we would like to complete this matrix.
(This was basically the task of the famous netflix prize.) In practice there are only a few types of significantly different client behavior which results
in the matrix being essentially of low rank. In abstract terms, we are given a matrix like this one
\[
\left(\begin{matrix} ? & 10 & ? & 2 & ? & ?\\
3 & ? & ? & ? & 3 & ?\\
? & ? & 14 & ? & ? &  14\\
? & 15 & 6 & ? & ? & ? \\
6 & ? & 4 & ? & 6 & 4\\
\end{matrix}\right),
\]
and the problem is to replace the question marks with numbers making the whole matrix being of low rank. 

Let $\Omega \subset [n_1] \times [n_2]$ of size $m$ be the location set of the known entries, where $[n] := \{1,2,\hdots,n\}$. Let $P_\Omega(X) \in \C^m$ be the restriction of a matrix $X \in \C^{n_1 \times n_2}$ 
to its entries in the set $\Omega$. The measurements in the matrix completion setup can then be written as $y = P_\Omega(X)$.
Compared to the subgaussian random measurement maps, $P_\Omega$ as a coordinate projection can be considered a structured measurement map. 
Later on, we will randomize this map by considering random subsets $\Omega$, and thereby obtain a structured random map.

Matrix completion, i.e., considering maps of the form $P_\Omega$, has some limitations in terms of low rank recovery.
Consider the rank-one matrix $X=e_j e_k^*$, where $e_j$ and $e_k$ are the $j$-th and $k$-th
canonical unit vectors in $\C^{n_1}$ and $\C^{n_2}$, respectively. Then $X$ is also one-sparse and if $(j,k) \notin \Omega$ then $P_\Omega(X) = 0$. Clearly, any reasonable
algorithm would recover the zero matrix from the zero measurement vector, so that $X$ cannot be recovered although it is of rank one. 
For this reason, $P_\Omega$ will not satisfy the rank restricted isometry property of any order. This means that we have to impose further conditions on the matrix 
apart from being low rank in order to guarantee recovery. 

It is natural to impose that the singular vectors of the matrix $X$ are incoherent with respect to the canonical basis.
In order to make this precise, for a subspace $U$ of $\C^n$ of dimension $r$ we introduce the coherence of $U$ as
\[
\mu(U) =\frac{n}{r} \max_{j=1,\hdots,n} \|P_U \mathbf{e}_j\|_2^2,
\]
where $P_U$ denotes the orthogonal projection onto $U$ and the $\mathbf{e}_j$ are the canonical unit vectors in $\C^n$. If $U$ contains a canonical unit vector, say $\mathbf{e}_j$,
then $P_U \mathbf{e}_j= \mathbf{e}_j$ and the coherence takes the maximal value $\mu(U) = n/r$.  The other extreme case is given for example by a space  
 $U$ spanned by $r$ orthonormal vectors  $\mathbf{u}_k$ maximally incoherent with respect to the canonical basis, i.e., $|\langle \mathbf{u}_k,\mathbf{e}_j\rangle| = 1/\sqrt{n}$. In this case, the matrix $R_U$ with rows $\mathbf{u}_k$ has all entries of modulus $1/\sqrt{n}$. Now $P_U=R_U^* R_U$, so $\|P_U \mathbf{e}_j\|_2=\|R_U \mathbf{e}_j\|_2 =\sqrt{\tfrac{r}{n}}$, and $\mu(U)$ takes the minimal 
value $\mu(U) = 1$.

When restricting to low rank 
matrices $X$ whose right and left singular vectors span incoherent subspaces in the sense that their coherence is small, it was shown by Cand{\`e}s and Recht in \cite{care09} 
that nuclear norm minimization is able to recover $X$ from $P_\Omega(X)$ from a random choice of $\Omega$ with high probability provided that enough measurements
are taken. The following statement \cite{gr09-2,re12} is a refinement of the first results in \cite{care09,cata10-2}.

\begin{theorem} Let $X \in \C^{n_1 \times n_2}$ of rank $r$ with reduced singular value decomposition $U \Sigma V^*$ with $U \in \C^{n_1 \times r}, \Sigma \in \C^{r \times r}$ and
$V \in \C^{n_2 \times r}$. Assume that the row and column spaces of $X$ satisfy $\mu(U), \mu(V) \leq \mu_0$ for some $\mu_0 \geq 1$, 
where with abuse of notation $U$ and $V$ also denote the 
subspaces spanned by the left and right singular vectors, respectively. Moreover, assume that
\[
\max_{j,k} |(U V^*)_{j,k}| \leq \mu_1 \sqrt{\frac{r}{n_1 n_2}}
\]
for some $\mu_1 \geq 1$. Let the entries of $\Omega \subset [n_1] \times [n_2]$ be sampled independently and uniformly at random. If 
\begin{equation}\label{bound:matrix:completion}
m \geq C \max\{ \mu_1^2,\mu_0\} r (n_1 + n_2) \log^2( n_1 + n_2)
\end{equation}
then $X$ is uniquely recovered from $P_\Omega (X)$ via nuclear norm minimization with probability at least \HR{$1-(n_1+n_2)^{-2}$}.
\end{theorem}
The bound \eqref{bound:matrix:completion} on the number of measurements is almost optimal in the sense that one can derive 
a lower bound where only the exponent $2$ at the log-factor is replaced by $1$. Moreover, note that the bound on $m$ requires $\mu_0$ to be small, which excludes
that the pathological sparse rank-one matrices $\mathbf{e}_j \mathbf{e}_k^*$ are recovered from fewer than $n_1 n_2$ measurements.

{\bf Random measurements with respect to an operator basis.} Motivated by problems from quantum tomography, Gross \cite{gr09-2} generalized the matrix completion setup
to the following scenario. Let $B_j \in \C^{n_1 \times n_2}$, $j=1,\hdots,n_1 n_2$, be an orthonormal basis with respect to the Frobenius inner product, i.e.,
$\langle B_j, B_k \rangle_F = \operatorname{tr}(B_j B_k^*) = \delta_{jk}$. Measurements of a matrix $X$ are taken with respect to this basis, i.e.,
\[
y_\ell = \langle X, B_{j_\ell}\rangle, \quad \ell=1,\hdots,m,
\]
for some $j_\ell \in [n_1 n_2]$. In this context, we define again a coherence parameter $\mu$, this time for the basis $\{B_j\}$, as
\[
\mu := \frac{n_1+n_2}{2} \max_{j=1,\hdots,n_1n_2} \| B_j\|_{2 \to 2}^2. 
\] 
The intuition is that sampling with respect to an operator basis having a small coherence parameter $\mu$ preserves information about low rank
matrices, and works well even for the pathological example $X = \mathbf{e}_j \mathbf{e}_k$. In the symmetric case $n_1=n_2 =n$, the optimal parameter $\mu = 1$ is taken for an operator basis $\{B_j\}$
for which $\sqrt{n} B_j$ is unitary for each $j \in [n^2]$. An example for such an operator basis is provided by the Pauli matrices, see \cite{gr09-2}, 
which are important in quantum mechanics. Another example is formed by time-frequency shift operators.
The following result from \cite{gr09-2} 
establishes low rank matrix recovery with respect to randomly chosen coefficients with respect to the orthonormal basis.
\begin{theorem}\label{thm:OpBasis} Let $\{B_j\}_{j=1}^{n_1n_2}$ be an operator basis with coherence $\mu \geq 1$. 
Let $X \in \R^{n_1 \times n_2}$ be of rank $r$ and  $\Omega \subset [n_1 n_2]$ be a subset of size $m$ which is chosen uniformly at random.
If, for $\varepsilon > 0$,
\[
m \geq C \mu r (n_1 + n_2) \log(n_1 n_2), 
\]
then $X$ is uniquely recovered from the samples $y_j = \langle X, B_j\rangle$, $j \in \Omega$, via nuclear norm minimization with \HR{probability at least $1-(n_1+n_2)^{-2}$}.
\end{theorem}
Another version of this result, which includes the matrix completion setup as a special case 
by considering the operator basis $\mathbf{e}_j \mathbf{e_k}, j=1,\hdots,n_1, k=1,\hdots,n_2$, can be found in \cite{gr09-2}.
Moreover, for the map of Theorem~\ref{thm:OpBasis}, also the rank-restricted isometry property \eqref{rankRIP} holds with \HR{probability at least $1-\gamma$} under the condition
\[
\HR{m \geq C\delta^{-2} \mu r (n_1 + n_2) \max\{ \log^6(n_1 + n_2), \log(\gamma^{-1})\}},
\]
see \cite{li11}. This implies uniformity of reconstruction as well as stability under passing to approximately low rank matrices and adding noise on the measurements. 
(Recall however, that for the matrix completion map the restricted isometry property fails.)

{\bf Fourier type measurements.} Let us now describe a structured measurement map connected to the Fourier transform for which
the rank restricted isometry holds. This map is the concatenation of random sign flips and a randomly subsampled two-dimensional Fourier transform.

In mathematical terms, for an $n_1 \times n_2$ matrix ${\mathcal E}$ with independent $\pm 1$ Rademacher entries $\epsilon_{j,k}$, we denote by
${\mathcal D}_{\mathcal E} : \C^{n_1 \times n_2} \to \C^{n_1 \times n_2}$ the Hadamard multiplication with $\mathcal E$, i.e.,
\[
\left({\mathcal D}_{\mathcal E}(X) \right)_{j,k} = \epsilon_{j,k} X_{j,k}.
\]
In other words, ${\mathcal D}_{\mathcal E}$ performs independent random sign flips on all entries of a matrix. Moreover, let ${\mathcal F} : \C^{n_1 \times n_2} \to \C^{n_1 \times n_2}$
denote the two-dimensional Fourier transform, i.e.,
\[
({\mathcal F} X)_{j,k} = \sum_{r=1}^{n_1} \sum_{t=1}^{n_2} X_{r,t} e^{-2 \pi i (r j /n_1 + kt/n_2)}.
\]
Finally, for a set $\Omega \subset [n_1]\times[n_2]$ of size $m$ we let $P_\Omega : \C^{n_1 \times n_2} \to \C^m$ be the restriction operator $(P_\Omega X)_{j,k} = X_{j,k}$ for 
$(j,k) \in \Omega$. With these ingredients our measurement map ${\mathcal A} : \C^{n_1 \times n_2} \to \C^m$ can be written as
\[
{\mathcal A}(X) = \frac{1}{\sqrt{m}} P_\Omega({\mathcal F}({\mathcal D}_{\mathcal E}(X))),
\]
where $\Omega$ is chosen uniformly at random among all subsets of $[n_1] \times [n_2]$ of cardinality $m$. Exploiting the FFT, the map ${\mathcal A}$ can be applied fast.

It is argued in the introduction of \cite{forawa10} (but details are not worked out), that ${\mathcal A}$ possesses the rank restricted isometry property \eqref{rankRIP} with high probability provided that
\[
m \geq C \delta^{-2} r (n_1 + n_2) \log^4(n_1 n_2).
\] 
It is interesting to note that this result follows from a combination of several facts: $\frac{1}{\sqrt{m}}P_{\Omega}{\mathcal F}$ satisfies the standard restricted isometry property \eqref{eq:RIP}
with high probability. Together with the main result in \cite{krwa11} relating Johnson-Lindenstrauss embeddings and the restricted isometry property if follows that 
$\frac{1}{\sqrt{m}} P_{\Omega} {\mathcal F}({\mathcal D}_{\mathcal E}(X))$ satisfies a certain concentration inequality which can then be used along with 
$\epsilon$-net arguments \cite{capl11} in order to establish the rank restricted isometry property.

\section{Phaseless Estimation} \label{sec:PE}
While the signal model used for low-rank matrix recovery and compressed sensing are considerably different, phaseless estimation problems consider structurally different, non-linear measurements. Namely, of each linear measurement, only the (squared) modulus is observed, and the phase is lost. Such a measurement setup arises in various applications such as X-ray crystallography and diffraction imaging. Losing the phase here corresponds to observing only the intensity of the measurements.

In more mathematical terms, the measurements take the form
\[
y = {\mathcal A}(x),
\]
where the non-linear map $\mathcal{A}:\C^N \text{ or } \R^N \rightarrow \R^\MM$ is given by $({\mathcal A} x)_j= |\langle a_j, x \rangle|^2$, where $\{a_i\}_{i=1}^\MM$ are given measurement vectors. Note that in this setup, the two cases of signal entries in $\R$ and $\C$ are structurally different as in the real case, the phase corresponds to the sign and hence only allows for the two different values $\pm 1$ whereas in the complex case, there are infinitely many possible phases. 

In the motivating application scenarios, the natural measurement vectors $a_i$ are again discrete Fourier basis vectors. As this obviously does not suffice for recovery, not even in certain cases or under additional assumptions, one often considers, in addition, phaseless coordinate measurements. That is, $a_i=e_i$, where $e_i$ is the $i$-th standard basis vector. This extended set of measurements is, in general, also known to not suffice to ensure uniqueness of the solution. However, as additional measurements are typically not available, there have been numerous works in the physics and optimization literature proposing efficient algorithms, showing empirically that they often yield good solutions, and deriving run time guarantees, see for example \cite{fi82}, \cite{bacolu03}, and the references therein.

We will again take a different viewpoint here. Namely, the goal will be to design measurements such that they allow for guaranteed recovery of the signal. As the measurements are always invariant under multiplication by a phase factor, all one can hope for, however, is recovery up to a global phase.  

A first natural question to ask concerns the minimal number of measurements such that the measurement map is injective. A number of works have been studying this question, often combining methods from frame theory and algebraic geometry. 
Typically, they show injectivity for {\em generic} sets of measurement vectors. Here generic means that the set of measurement vector configurations that yield an injective map $\mathcal A$ (up to a global phase) is open and dense in the set of frames. Recall that a frame is a set $\{f_j\}_{j=1}^\MM$ of vectors in $\R^N$ or $\C^N$ such that there exists $A,B>0$ such that for all 
$x$ in $\R^N$ or $\C^N$ it holds that $A\|x\|_2^2\leq \sum_{j=1}^\MM |\langle f_j, x\rangle|^2 \leq B\|x\|_2^2$.

In the real case, the question of injectivity is completely answered in the following result from \cite{bacaed06-1}.
\begin{theorem}
 For a generic frame $\{a_j\}_{j=1}^\MM \subset \R^N$, the phaseless measurement operator $\mathcal A$ corresponding to the measurement vectors $a_j$ is injective provided $\MM\geq 2N-1$. 
 On the other hand, no set of $\MM<2N-1$ measurement vectors in $\R^N$ yields an injective operator $\mathcal A$.
\end{theorem}
The complex case is less understood. It is known that $\MM=4N-4$ generic measurement vectors suffice to achieve injectivity \cite{coedhevi13}, and that $4N-o(N)$ measurements are necessary \cite{hemawo13}, see also \cite{movo13}. While it has been conjectured that $\MM=4N-4$ measurements are also necessary \cite{MixonBlog}, this question is currently open. 

While injectivity is certainly a useful indicator for when the phaseless estimation problem has a chance of being solved, it does not imply anything about the existence of tractable recovery algorithms nor about the conditioning (and hence the possibility of an efficient inversion in case of noisy measurements). A number of works addressing such issues are based on the observation that the measurement information can be expressed in matrix form. Namely, for $X=xx^*\in \K^{N\times N}$ and $A_i=a_i a_i^*\in \K^{N\times N}$, the constraint $|\langle a_i, x\rangle|^2=y_i$ can be reexpressed as $\langle A_i, X\rangle =y_i$, where one considers the Hilbert-Schmidt inner product of matrices $\langle B, C\rangle = \tr A^* B$. In this formulation, the constraints are again linear in $A$. Thus for on the order of $\MM=\tfrac{N(N+1)}{2}$ (in the real case) 
or $\MM=N^2$  suitably chosen measurements, one can directly solve for the $\tfrac{N(N+1)}{2}$ or $N^2$ entries of the matrix $X$ \cite{babocaed09} (the reduced 
number of entries in the real case stems from the fact that then $X$ is symmetric). The global phase ambiguity mentioned above is also incorporated in this formulation, as changing $x$ by a global phase does not change $X$ (and $x$ can be determined from $X$ up to a global phase).

A number of measurements quadratic in the signal dimension, however, is quite far from the injectivity benchmark of linear scaling in $N$. 
One may hope for a significantly smaller sufficient number of measurements because $X$ is of rank one. Hence while for a subquadratic number of measurements, the matrix reformulation admits additional (matrix) solutions, $X$ is definitely the one of smallest rank. Thus the problem boils down to recovering a low-rank matrix from linear measurements. In contrast to the setup derived in Section~\ref{sec:MR}, however, the measurements correspond to inner products with rank one matrices and do not satisfy any of the conditions required in the results presented there. As it turns out, however, the same strategy of convex relaxation still works. As the matrices considered here are all positive semi-definite, the nuclear norm of the matrix is just the trace. The resulting algorithm, coined PhaseLift, is introduced in \cite{caelstvoXX} as the following 
minimization problem.
\begin{equation}
 \widehat{X}= \argmin\limits_{X \succcurlyeq 0, {\mathcal A}X=y} \tr(X) \tag{PL} \label{eq:PL}
\end{equation}
Here $X\succcurlyeq 0$ means that $X$ is a positive semi-definite matrix. 
Similar to \eqref{eq:l1}, the formulation can also be adapted to noisy measurements. As noted in \cite{caliXX} and \cite{dehaXX}, with high probability there is just a single positive semi-definite matrix $X$ that satisfies the measurement constraints. Hence, the problem becomes a feasibility problem rather than a convex optimization problem.

PhaseLift is a tractable algorithm, but not comparable in efficiency to the heuristic algorithms mentioned above. Nevertheless, PhaseLift is considered a breakthrough for the analysis of the phaseless estimation problem, as for the first time, recovery guarantees could be established. The first scenario 
considered was that the measurements are chosen to be Gaussian vectors or uniformly distributed on the unit sphere. As the measurement vector and hence also its length is known, these two measurement setups are equivalent when no noise is considered. In \cite{castvoXX}, non-uniform recovery guarantees were established with high probability for a number of measurements on the order of $N\log(N)$. In \cite{caliXX}, these results were refined to yield uniform recovery guarantees for a number of measurements scaling linearly in $N$. As obviously no recovery is possible for less measurements than the dimension, these embedding dimensions are optimal up to an absolute constant. The result from \cite{caliXX} reads as follows.
\begin{theorem}
 Assume that the number of measurements satisfies $\MM\geq c_0 N$, where $c_0$ is a sufficiently large constant. For $i=1,\dots, \MM$, define $y_i(x)=|\langle a_i, x\rangle|^2$, 
 where the $a_i$ are independent standard Gaussian vectors. Then with probability \HR{at least $1-e^{-c_1 \MM}$} on the draw of the measurement vectors, it holds that for all $x$, 
 the solution to the minimization problem \eqref{eq:PL} exactly agrees with the signal $x$. The recovery is uniform in the sense that the same draw guarantees recovery for all $x$ from $y(x)$ simultaneously.
\end{theorem}
Note that in contrast to compressed sensing, a direct generalization from Gaussian vectors to vectors with subgaussian entries is not possible. Namely, for Rademacher measurement vectors recovery cannot be possible, as each of the standard unit vectors yields the exact same phaseless measurements, so they cannot be distinguished. Thus one needs an additional condition on the small ball probabilities of the entries of the measurement vectors. We refer to \cite{elmeXX} for details, where the authors do not consider a specific algorithm, but rather derive stability in the sense that signals that significantly differ also yield measurements that are not too close. Moreover, \cite{vo13} considers recovery via PhaseLift from measurements with random unitary matrices.

A first attempt to derive recovery guarantees for different, more efficient algorithms was the polarization algorithm provided in \cite{albafimiXX}. Later, in \cite{nejasa13}, the authors consider an alternating minimization algorithm, which is inspired by the heuristic algorithms mentioned above. Both these papers provide recovery guarantees for independent Gaussian measurement vectors.

The first paper that derived theoretical guarantees for structured measurements in phase retrieval was \cite{bachmi13}. Their work is motivated by applications in diffraction imaging. As mentioned above, measurements in this setup are absolute values of Fourier coefficients, which, a priori, do not suffice to recover the signal. As suggested in \cite{fa12}, however, one can introduce masks into the measurement setup, which have the effect that only parts of the object are illuminated. By varying the mask, one can obtain multiple images and thus more information in total. With this modification, the above limitations do not apply, so recovery of the signal is possible. In \cite{bachmi13}, the authors derive recovery guarantees for the polarization algorithm introduced in \cite{albafimiXX} and masked Fourier measurements. The number of measurements they require is of order $N\log(N)$.

For PhaseLift, the first structured measurement setup is provided in \cite{grkrku13}. The paper considers measurements selected uniformly at random from 
spherical designs; the number of measurements necessary to guarantee recovery depends on the order of the design. Basically at the same time, \cite{caliso13} considers PhaseLift for masked Fourier measurements and derives recovery guarantees for a number of measurements on the order of $N \log^4(N)$, that is, $\log^4 N$ masks. Subsequently, these guarantees have been  improved to require only $\log^2 N$ masks \cite{grkrku14}.

In many works on the phaseless estimation problem, there have been attempts to incorporate sparsity assumptions 
into the problems to reduce the number of required measurements, see \cite{elseshsz11,shbeel14} for first algorithmic
and applied contributions.
Refined recovery guarantees for $k$-sparse signals have first been provided in \cite{livo12}. 
They provide recovery guarantees for a modification of PhaseLift and a number of measurements on the order of $k^2 \log(N)$. 
As they show, the quadratic dependence on $k$ is necessary in their algorithmic setup, see also \cite{elfajaoy12} for similar lower bounds.
As shown in \cite{elmeXX}, the number of Gaussian measurements required for stability is of order $k \log(N)$, hence considerably smaller. 
Thus for sparse signals, the PhaseLift approach cannot work with optimal embedding dimensions. Based on this work, \cite{ehfosi13} provides recovery guarantees for sparse vectors with an additional decay condition using a greedy algorithm for a number of measurements on the order of $k \log(N)$.

\section{Mathematical proof techniques}\label{sec:MM}
Due to the comprehensive nature of this survey article, we cannot give a self-contained mathematical exposition of the proof techniques. In the following, however, we provide some key words and references for the different classes of results presented in this article.

For unstructured random matrices/maps, the proof of the restricted isometry property is by-now rather standard: Using Bernstein's inequality one establishes
a concentration inequality for $\|A x\|_2^2$ for a fixed $x$. 
Then one covers the $\ell_2$-sphere restricted to the sparse vectors/low rank matrices with a relatively dense but finite collection of vectors, a so-called $\epsilon$-net, takes
a union bound of the concentration inequality over the net, and extends the resulting estimate to the whole infinite set of interest by a bootstrapping argument 
\cite{badadewa08,capl11,fora13,mepato09}.

As one can imagine, proving recovery guarantees for structured random measurement matrices/maps is considerably harder because these matrices contain 
much less randomness. For instance, in order to establish the restricted isometry property for random partial Fourier matrices, one first applies
symmetrization \cite{fora13,leta91}, followed by the Dudley inequality for the expected supremum of a subgaussian processes which leads to an integral over covering
numbers with respect to a certain metric. Then one uses techniques such as the Maurey Lemma \cite{ca85,krmera12} in order to bound the resulting covering numbers.

The Dudley inequality just mentioned is itself proved using the so-called chaining technique. Talagrand has developed a much more general theory of generic chaining \cite{ta05-2},
and in fact, the bound of the restricted isometry property for partial random circulant matrices (Theorem~\ref{thm:KMR}) is based on a new generic chaining bound
for certain chaos processes \cite{krmera12}, which then again requires to obtain bounds for associated covering numbers.

One ingredient for establishing nonuniform recovery guarantees are condition number bounds for a single submatrix of the measurement matrix corresponding
to the columns indexed by the support set. These often require to estimate the operator norm of a sum of independent random matrices. Traditionally, noncommutative
Khintchine  inequalities \cite{bu01,lu86-1,ra10} were used for this purpose, but recently, Tropp \cite{tr12} developed extensions of 
many classical deviation inequalities such as Bernstein's inequality to the matrix setting, which are much simpler to use. In the case of partial random circulant matrices \cite{ra09,ra10}
and similar setups \cite{pfra10} one ends up with a double sum of random matrices, a so-called second-order matrix-valued chaos for which an extension of Khintchine's inequality
can be used to obtain operator norm bounds \cite{ra09,ra10}. Alternatively, the so-called trace method leads to combinatorial estimates \cite{carota06,ra07,pfra10,ja13}.

In addition to establishing condition number bounds for the submatrix of the measurement matrix corresponding to the support of the sparse vector, one essentially needs to show
that this submatrix behaves well with respect to the columns outside the support. This task is often the harder part of the analysis in the nonuniform setting. In the initial contribution \cite{carota06}, it was
established by following complicated combinatorial arguments, see also \cite{ra07,pfra10,hurast12}. A more elegant approach -- called golfing scheme -- was developed by
Gross \cite{gr09-2} in the context of matrix completion. It proceeds via introducing an artificial iteration by partitioning the matrix into smaller blocks of rows. 
However, so far it seems that this approach is restricted to matrices
with stochastically independent rows. We refer to \cite{capl11-1,fora13,gr09-2,re12} for details.

For more information on corresponding probabilistic techniques and for background information on nonasymptotic random matrix theory in general, 
we refer to \cite{chgulepaXX,fora13,ra10,leta91,elku12,ta05-2,ve12}.

\subsection*{Acknowledgements}
F. Krahmer acknowledges support by the German Federal Ministry of Education and Reseach (BMBF) through the cooperative research project ZeMat. H. Rauhut acknowledges support by the European Research Council through the Starting Grant 258926 (SPALORA).

\vspace{\baselineskip}


\end{document}